\documentclass[a4paper]{jpconf}
\usepackage{graphicx}
\begin{document}
\title{Ultralight dark matter: constraints from gravitational waves and other astrophysical observations}

\author{Tanmay Kumar Poddar}

\address{Theoretical Physics Division, Physical Research Laboratory, Ahmedabad-380009, India}

\ead{tanmay@prl.res.in}

\begin{abstract}
The orbital period loss of the compact binary systems is the first indirect evidence of gravitational waves which agrees well with Einstein's general theory of relativity to a very good accuracy. However, there is less than one percent uncertainty in the measurement of orbital period loss from the general reltivistic prediction. Perihelion precession of planets, Gravitational light bending and Shapiro delay are three other successful tests of general relativity theory. Though there are uncertainties in the measurements of those observations from the general reltivistic predictions as well. To resolve these uncertainties, we assume radiation of ultralight axions and light gauge boson particles of $L_i-L_j$ type from those systems which can be a possible candidate of fuzzy dark matter. In this article, we obtain bounds on new physics parameters from those astrophysical observations.
\end{abstract}

\section{Introduction}
The first indirect evidence of gravitational wave (GW) was observed from the orbital period loss of Hulse-Taylor (HT) compact binary system which agrees well with Einstein's general relativistic prediction with less than one percent uncertainty. However, the first direct evidence of GW was confirmed in 2015 from the merger of two steller mass black holes. The other successful tests of Einstein's GR theory are perihelion precession of Mercury, gravitational light bending and Shapiro delay. The observational values of those tests match with the GR values with the uncertainties of $\mathcal{O}(10^{-3})$ for perihelion precession, $\mathcal{O}(10^{-4})$ for light bending and $\mathcal{O}(10^{-5})$ for Shapiro delay. To fix these uncertainties, we assume there may be radiation of ultralight particles like axions and other light gauge bosons of $L_i-L_j$ type from those astrophysical systems which contribute to those uncertainties. These ultralight particles can be a promising candidate of fuzzy dark matter (FDM). The ultralight axions with mass $10^{-22}\rm{eV}$ can be a candidate of FDM if the axion decay constant is $\mathcal{O}(10^{17}\rm{GeV})$. The de Broglie wavelelngth of such ultralight particles is of the order of the size of a dwarf galaxy (1-2kpc). These ultralight dark matter can solve some of the small scale structure problem of the universe and also free from direct detection bounds.
\section{Constraints on ultralight axions from compact binary systems}
\begin{figure}[h]
\begin{center}
\includegraphics[width=20pc]{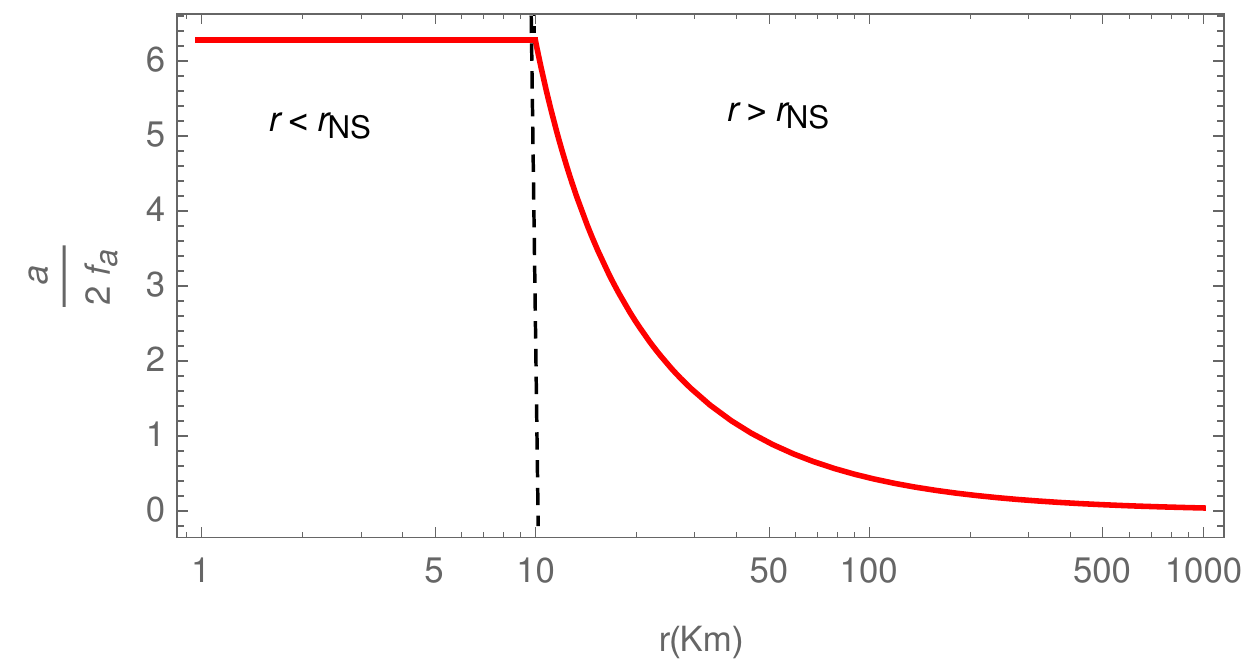}
\caption{\label{m1}Variation of axion field with distance for a compact object. }
\end{center}
\end{figure}
The typical value of the orbital frequency of the compact binary system is $\mathcal{O}(10^{-19}\rm{eV})$. Hence, if we assume particles with such ultralight mass or even less emit from the binary system then it can resolve the uncertainty in the measurement of orbital period loss compared with the GR prediction. We assume compact stars such as neutron stars (NS) and white dwarfs (WD) are immersed in a long range axionic potential and if the orbital frequency is greater than the mass of axion like particles (ALPs) then the compact system results a long range axion hair outside of the binary system. Here we assume interaction between axions with the quarks. In Table \ref{ex1} we consider four compact binary systems (two NS-NS and two NS-WD) and obtain bounds on axion decay constant $f_a$. The $\alpha$ denotes the ratio of axion mediated fifth force to the gravitational force. We obtain the stronger bound on $f_a$ is $\leq 10^{11}\rm{GeV}$ for axions of mass $\leq 10^{-19}\rm{eV}$ and the mass is in the ballpark of FDM. This implies if ALPs are FDM, then they do not couple with quarks. In Figure \ref{m1} we have shown the variation of axion field with the distance of a compact object (here NS) and in Figure \ref{m1} we have shown the variation of axion potential with the axion field for inside and outside of a compact object \cite{one}.
\begin{table}[h]
\caption{\label{ex1}Summary of the upper bounds on $f_a$ for four compact binary systems. These bounds are for ALPs with mass $m_a\leq 10^{-19}\rm{eV}$.}
\begin{center}
\begin{tabular}{llll}
\br
Compact binary system &$f_a$ (GeV) & $\alpha$\\
\mr
PSR J0348+0432  & $\leq 1.66\times 10^{11}$  & $\leq 5.73\times 10^{-10}$ \\
 PSR J0737-3039 & $\leq 9.69\times 10^{16}$  &$\leq 9.21\times 10^{-3}$ \\
 PSR J1738+0333 & $\leq2.03\times 10^{11}$  & $\leq 8.59\times 10^{-10}$\\
PSR B1913+16 &     $\leq2.07\times 10^{17}$  & $\leq 3.4\times 10^{-2}$ \\
\br
\end{tabular}
\end{center}
\end{table}
\section{Vector gauge boson radiation from compact binary systems in a gauged $L_\mu-L_\tau$ scenario}
\begin{figure}[h]
\begin{center}
\includegraphics[width=20pc]{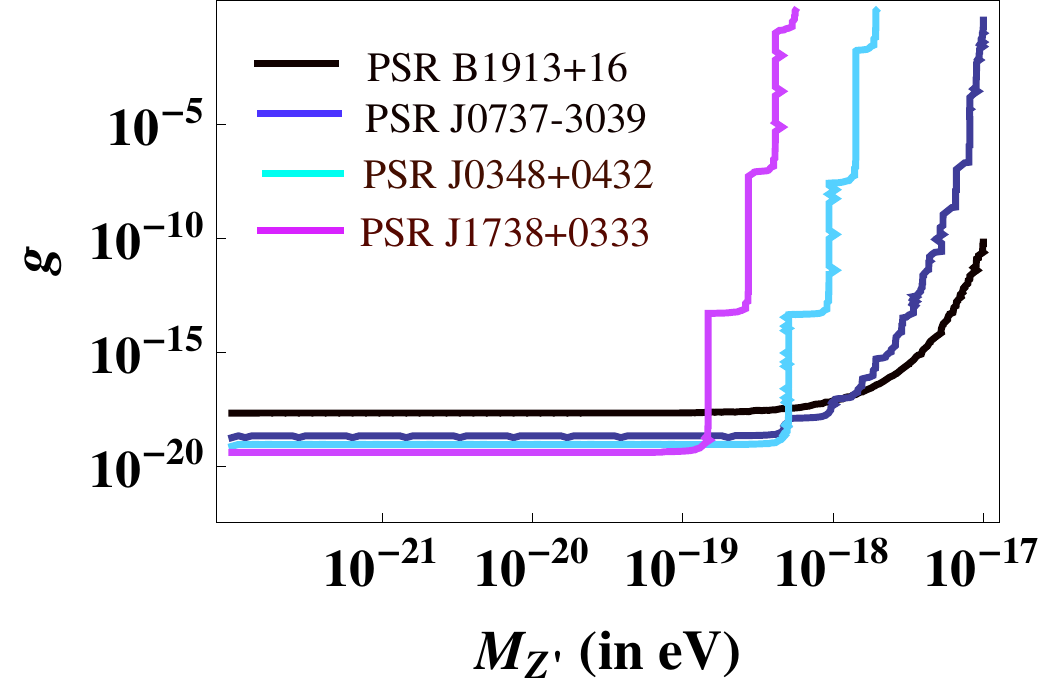}
\caption{\label{m2}Variation of gauge coupling with the gauge boson mass in a gauged $L_\mu-L_\tau$ scenario}
\end{center}
\end{figure}
Due to the presence of large fraction of muons $(N_\mu\sim 10^{55})$ in NS, compact binary systems like NS-NS and NS-WD can mediate long range $L\mu-L_\tau$ type of gauge force. The large chemical potential of degenerate electrons hinders muon decay in NS which results lots of muons in that compact object. Though WDs do not contain any muon charge. We assume $L_\mu-L_\tau$ type of gauge bosons can emit from the binary systems and can contribute to the uncertainty in the orbital period loss of the binary system compared with the GR prediction. In Table \ref{ex2} we obtain $L_\mu-L_\tau$ gauge couplings for four compact binary systems. The stronger bound on the gauge coupling is $g\leq 4.24\times 10^{-20}$ \cite{two}. The variation of $L_\mu-L_\tau$ gauge coupling with the gauge boson mass is shown in Figure \ref{m2}. 
\begin{table}[h]
\caption{\label{ex2}Summary of gauge couplings for four compact binary systems. The mass of the gauge boson is constrained by the orbital frequency of the binary systems which is $M_{Z^\prime}\leq 10^{-19}\rm{eV}$.}
\begin{center}
\begin{tabular}{llll}
\br
Compact binary system \hspace{0.5cm} & $g$(fifth force)\hspace{0.5cm} & $g$(orbital period decay)\\
\mr
PSR B1913+16  & $\leq 4.99\times 10^{-17}$  & $\leq 2.21\times 10^{-18}$ \\
PSR J0737-3039 & $\leq 4.58\times 10^{-17}$  &$ \leq 2.17\times 10^{-19}$\\
PSR J0348+0432 & $ -$  & $\leq 9.02\times 10^{-20}$ \\
PSR J1738+0333 & $ -$  &$ \leq 4.24\times 10^{-20}$\\
\br
\end{tabular}
\end{center}
\end{table}
\section{Constraints on long range force from perihelion precession of planets in a gauged $L_e-L_{\mu,\tau}$ scenario}
\begin{figure}[h]
\begin{center}
\includegraphics[width=20pc]{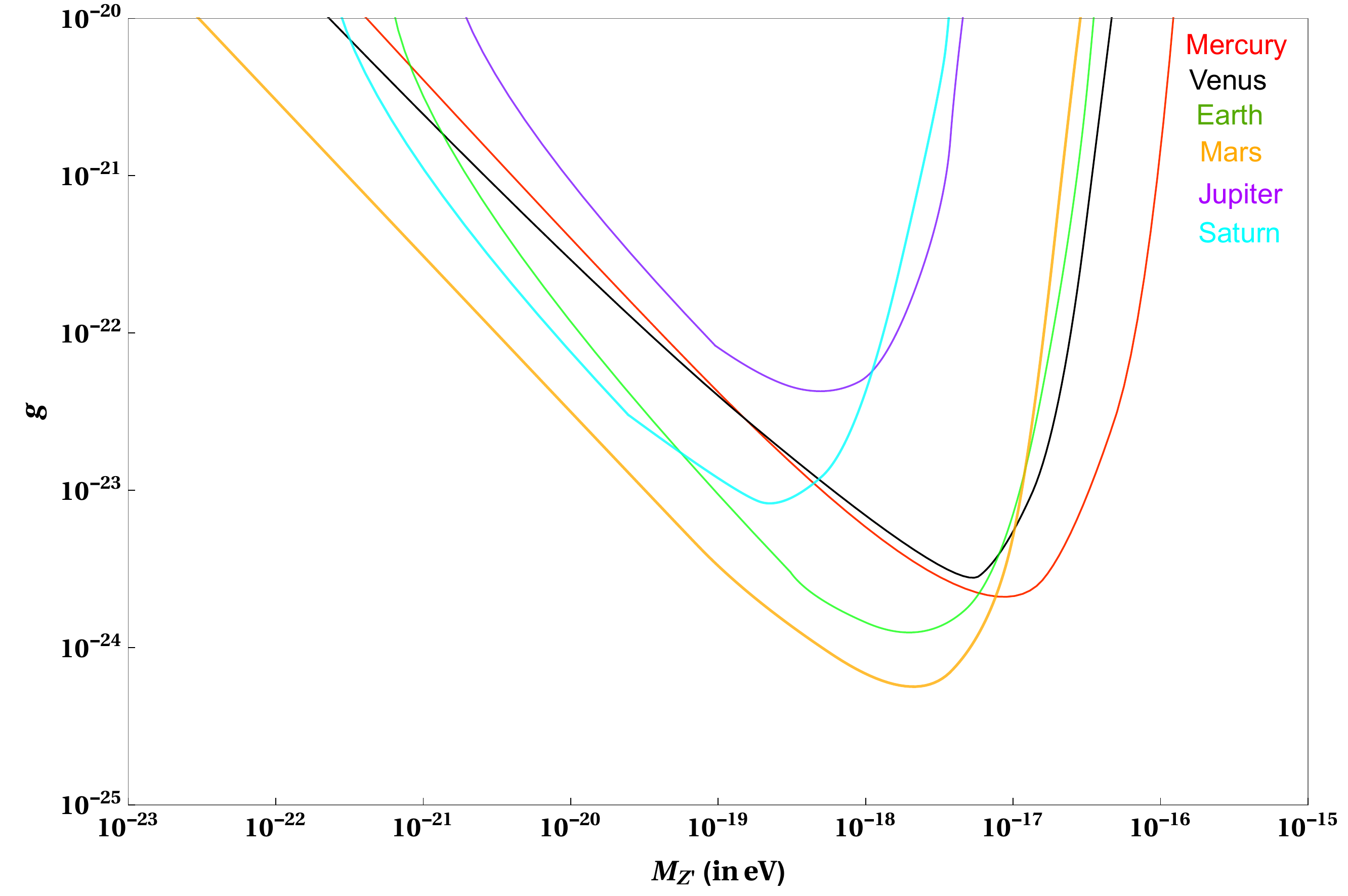}
\caption{\label{m3}Variation of gauge coupling with the gauge boson mass in a gauged $L_e-L_{\mu,\tau}$ scenario .}
\end{center}
\end{figure}
The presence of electrons in planets and Sun results a long range Yukawa mediated $L_e-L_{\mu,\tau}$ gauge force between them. We assume that the gauge bosons of $L_e-L_{\mu,\tau}$ type mediate between the celestial bodies can resolve the uncertainties in the perihelion precession of planets compared to the GR prediction. We obtain the upper bounds on the gauge coupling of $L_e-L_{\mu,\tau}$ type for all the planets upto Saturn and the planet Mars gives the stronger bound as $g\leq 3.506\times 10^{-25}$. The mass of the gauge boson is constrained from the distance between the planet and the Sun \cite{three}. The variation of $L_e-L_{\mu,\tau}$ gauge coupling with the gauge boson mass is shown in Figure \ref{m3}.
\section{Probing the angle of birefringence due to long range axion hair from pulsars }
\begin{figure}[h]
\begin{center}
\includegraphics[width=20pc]{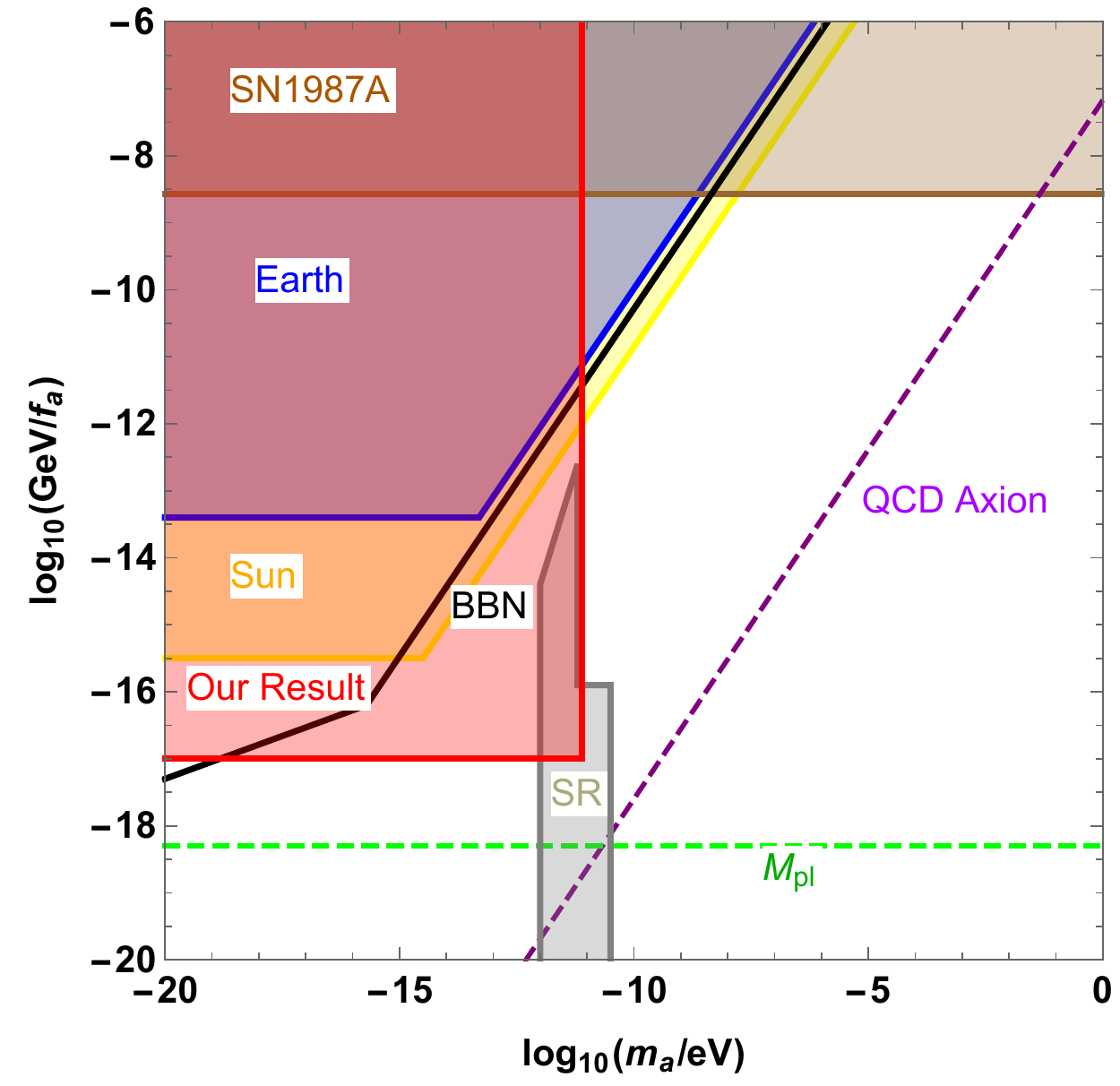}
\caption{\label{m4}Variation of axion decay constant with axion mass. We can probe the red shaded region. The other shaded regions are excluded from different astrophysical experiments.}
\end{center}
\end{figure}
Earlier we have discussed that rotating neutron stars or pulsars can be a source of axions and for an isolated neutron star, the mass of the axion is constrained by the radius of the compact object which yields $m_a\leq 10^{-11}\rm{eV}$. Also pulsars can emit electromagnetic radiation. We consider the long range axion hair emitted from pulsar can rotate the polarization of the electromagnetic radiation and the result is within the accuracy of measuring the linear polarization angle of pulsar light which is $\Delta\theta =0.42^\circ$. Our result is true for axions with $f_a\leq10^{17}\rm{GeV}$ \cite{four}. The variation of axion decay constant with axion mass is shown in Figure \ref{m4}.
\section{Constraints on axionic fuzzy dark matter from light bending and Shapiro time delay}
\begin{figure}[h]
\begin{center}
\includegraphics[width=20pc]{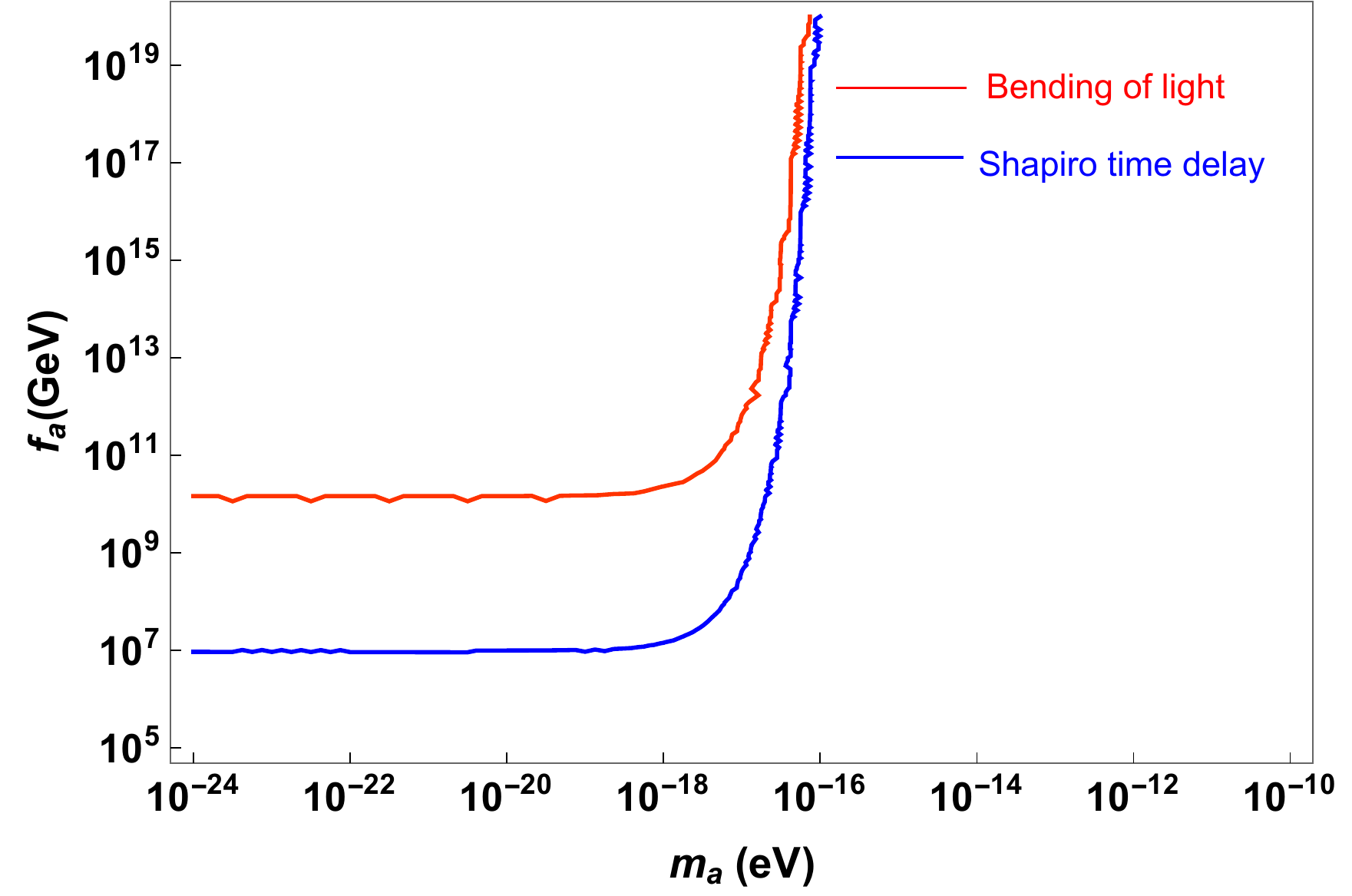}
\caption{\label{m5}Variation of axion decay constant with axion mass for light bending and Shapiro delay.}
\end{center}
\end{figure}
In this section we consider two observations namely gravitational light bending and Shapiro time delay to constrain axionic fuzzy dark matter. The celestial objects like Earth and Sun can have a large number of axions and its radiation from those systems can contribute to the uncertainties in the measurements of those observations compared to the GR prediction. The mass of the axion is constrained from the distance between Earth and Sun which yields $m_a\leq 10^{-18}\rm{eV}$. In Table \ref{ex3} we obtain bounds on the axion decay constant and the Shapiro time delay puts the stronger bound $(f_a\leq 9.85\times 10^6\rm{GeV})$ \cite{five}. The results demand if ALPs are FDM then they do not couple with quarks. The $\alpha$ in the Table denotes the ratio of the axion mediated fifth force to the gravitational force. The variation of axion decay constant with the axion mass for light bending and Shapiro delay is shown in Figure \ref{m5}.
\begin{table}[h]
\caption{\label{ex3}Summary of axion decay constants for light bending and Shapiro time delay. The mass of the axion is constrained from the distance between Earth and Sun which yields $m_a\leq 10^{-18}\rm{eV}$.}
\begin{center}
\begin{tabular}{llll}
\br
Experiments & axion decay constant ($f_a$) & $\alpha$  \\
\mr
Light bending  &  $ \leq 1.58\times 10^{10}\rm{GeV}$ & $\leq10^{-2}$\\
 Shapiro time delay & $ \leq 9.85\times 10^{6}\rm{GeV}$ & $\leq4.12\times 10^{-9}$\\
\br
\end{tabular}
\end{center}
\end{table}
\section{Conclusions}
The precision measurements of GW and other astrophysical experiments demand a possibility of radiation of light particles like axions, gauge bosons etc. In this article, we have discussed the bounds on the new physics parameters. Physics of those light particles is interesting as it can be a possible dark matter candidate. One can also probe $U(1)_{L_i-L_j}$ scenarios from those above experiments.

\end{document}